\documentclass[prd,a4paper]{revtex4}

\usepackage{graphicx}
\usepackage{psfrag}
\usepackage{inputenc}
\usepackage{amsmath}
\usepackage{amssymb}
\usepackage{subfigure}

\inputencoding{latin9}

\newcommand{\nc}{\newcommand}

\def\gsim{\mathrel{\raise.3ex\hbox{$>$\kern-.75em\lower1ex\hbox{$\sim$}}}}

\renewcommand{\thesubfigure}{\thefigure.(\alph{subfigure})}
\makeatletter
\renewcommand{\p@subfigure}{}
\renewcommand{\@thesubfigure}{\thesubfigure\hskip\subfiglabelskip}
\makeatother

\nc{\be}[1]{\begin{equation}\mbox{$\label{#1}$}}
\nc{\bea}[1]{\begin{eqnarray} \mbox{$\label{#1}$}}
\nc{\Section}[2]{\section{#2}\label{#1}}
\nc{\Bibitem}[1]{\bibitem{#1}}
\nc{\Label}[1]{\label{#1}}
\nc{\ie}{{\em i.e.\ }}
\nc{\eg}{{\em e.g.\ }}
\nc{\eea}{\end{eqnarray}}
\nc{\ee}{\end{equation}}
\nc{\w}{\omega}
\begin{document}

\title{Suppressing the large scale curvature perturbation by interacting fluids}

\author{T. Multam\"aki}
\thanks{tuomul@utu.fi}
\affiliation{Department of Physics,
University of Turku, FIN-20014 Turku, FINLAND}
\author{J. Sainio}
\thanks{jtksai@utu.fi}
\affiliation{Department of Physics,
University of Turku, FIN-20014 Turku, FINLAND}
\author{I. Vilja}
\thanks{vilja@utu.fi}
\affiliation{Department of Physics,
University of Turku, FIN-20014 Turku, FINLAND}
\date{\today}

\begin{abstract}
The large-scale dynamics of a two-fluid system with a time dependent interaction
is studied analytically and numerically. 
We show how a rapid transition can significantly suppress the large-scale curvature
perturbation and present approximative formulae for estimating the effect. By comparing to 
numerical results, we study the applicability of the approximation and find good
agreement with exact calculations.
\end{abstract}

\maketitle

\section{Introduction}

There are many instances in cosmology, when the large-scale
 evolution of perturbations \cite{Mukhanov:1990me} in a multi-component system
 needs to be solved \cite{Wands:2000dp}. 
The use of curvature and 
entropy perturbations has become a \textsl{de facto} standard in this formalism,
pioneered by Kodama and Sasaki \cite{Kodama:1985bj}.
In a multi-component system, interactions between fluids are an important aspect
in determining the evolution of the curvature perturbation \cite{Malik:2004tf}.
The importance of such systems is evident: interacting fluids are the cornerstone of many
important cosmological scenarios, such as reheating at the end of inflation
\cite{Albrecht:1982mp,Den:1984tn,Kripfganz:1985mn,Bastero-Gil:2002xr,Dvali:2003em,Kofman:2003nx} 
or the curvaton scenario \cite{Enqvist:2001zp,Lyth:2001nq,Moroi:2002rd,Moroi:2001ct,Lyth:2002my}.

The total curvature perturbation, $\zeta$, generally evolves in multi-component fluid whenever the 
non-adiabatic pressure is non-zero. In particular, if the fluids are interacting the 
total curvature perturbation changes.
Whether the evolution changes the amplitude of the large-scale curvature perturbation significantly or not,
depends on the exact nature of the system. Efficient amplification, or damping, of $\zeta$ on large scales 
by later evolution allows one to modify the previously produced perturbation spectrum, relaxing
constraints on the scale at which the perturbations were produced. A natural framework for such 
mechanisms is \eg within a traditional multiple inflationary scenario
\cite{minf} or a string landscape picture \cite{cliff, multiverse}.

The transfer of energy between fluids can be described by a number of ways: for example 
in \cite{Malik:2002jb}
a constant interaction between different components is assumed whereas authors of \cite{Lyth:2001nq} 
and \cite{Lyth:2002my} 
use the so-called sudden decay approximation instead. In this approximation, the 
two fluids evolve independently until energy is transferred rapidly from one fluid to the other
at a particular point in time. This method allows one to estimate the magnitude of the 
total curvature perturbation analytically. Here we relax these assumptions and 
allow for a time-dependent interaction while evolving the full large-scale perturbation
equations with no further approximations. Research in this direction has recently been presented in 
 \cite{Dvali:2003em} and \cite{Kofman:2003nx}, where locally different decay rates of the inflaton are 
generated by spatially varying reheating temperature and couplings.
Our approach is to study how turning on the interaction affects the perturbations and
the significance of the transition time scale. 
When the interaction is turned on rapidly so that it is well modeled by a step-function, 
we can proceed with analytical methods and derive a useful approximation to the numerical 
results.  

In the fluid approach with constant interaction term, the microphysics cannot be properly
accounted for as the fluid begins to decay simply when its decay width is of the order
of the Hubble rate, $\Gamma\sim H$. Such a simple description does not always properly reflect the
physical process, however, since there can be additional physical scales and processes involved 
that can play an important role. Cosmological processes involving different physical scales,
such as phase transitions, do not need to conform to the simple fluid description but are
better described by a time-dependent interaction.

In this paper we will consider time dependent interactions between different fluids. In section 
II we present the governing equations of perturbations in the Newtonian gauge. In the following section 
we compare the evolution of perturbations by means of numerical and analytical methods. We present
analytical formulae which give an accurate approximation of the behavior of curvature 
perturbation variables and study the validity of these approximations.
We end the paper with a short conclusion. Precise calculations of the evolution of the
curvature perturbation of the decaying fluid during transition are presented in the Appendix.

\section{Perturbation equations}

Following the notation of \cite{Mukhanov:1990me,Malik:2004tf}:
we consider linear scalar perturbations about a spatially flat Friedmann-Robertson-Walker -background in 
the Newtonian gauge:
\be{metric}
-ds^2=g_{\mu\nu}dx^{\mu}dx^{\nu}=(1+2\phi)dt^2-a(t)^2(1-2\psi)\delta_{ij}dx^idx^j,
\ee
where $a$ is the scale factor.
The background evolution is determined by the Einstein's equations, $G_{\mu\nu}=8\pi G T_{\mu\nu}$, and
the continuity equations of the individual fluids:
\be{fluidcont}
\dot{\rho}_{0(\alpha)}+3H(1+\w_{(\alpha)})\rho_{0(\alpha)}=Q_{\alpha},
\ee
where $\w_{(\alpha)}$ is the equation of state and $Q_{\alpha}$ describes the interactions between 
the different fluids. Total energy density is conserved, hence $\sum_{\alpha}Q_\alpha=0$.
Perturbing the covariant continuity equations, one finds the evolution equations of the perturbed energy and pressure densities
(on large scales) \cite{Malik:2004tf}:
\be{fluidpert}
{\delta\dot\rho}_{\alpha}+3H(\delta\rho_{(\alpha)}+\delta P_{(\alpha)})-3(\rho_{0(\alpha)}+P_{0(\alpha)})\dot{\psi} = Q_\alpha\phi + \delta Q_\alpha.
\ee
In addition, we have the $G^{0}_{0}$ component of perturbed Einstein's equations
\begin{equation} \label{psieq}
3H\big(\dot{\psi}+H\phi\big) = -4\pi G\delta\rho,
\end{equation}
where $\delta\rho = \sum_\alpha \delta\rho_\alpha$.
Note that for perfect fluids $\phi=\psi$ in the Newtonian gauge, and hence given the equations of state, $\w_{(\alpha)}$,
and the interactions between the fluids, $Q_\alpha$, one can evolve the individual
fluid perturbations along with the metric perturbation $\phi$.

The curvature perturbation of the fluid component $\alpha$ 
is defined to be the metric perturbation $\psi$ on uniform $\alpha$-fluid density hypersurfaces,
\begin{equation} \label{eq:fluid-curv}
\zeta_{(\alpha)} \equiv -\psi - H\frac{\delta \rho_{(\alpha)}}{\dot{\rho}_{0(\alpha)}} = -\psi - \frac{\delta \rho_{(\alpha)}}{\rho_{0(\alpha)}'},
\end{equation}
where comma represents derivative with respect to the number of e-folds $N$, 
$\ '\equiv d/dN\equiv d/d(\ln a)$.
The total curvature perturbation can be expressed as  sum of the individual curvature perturbations,
\begin{equation} \label{eq:fluid-curv-tot}
\zeta \equiv
\sum_\alpha \frac{\dot{\rho}_{0(\alpha)}}{\dot{\rho_0}}
\zeta_{(\alpha)},
\end{equation}
where $\rho_0\equiv \sum_\alpha\rho_{0(\alpha)}$.

From the definition of the curvature perturbations (\ref{eq:fluid-curv}) one discerns that if $\dot{\rho}_{0(\alpha)}=0$,
the variable $\zeta_{(\alpha)}$ becomes ill-defined. Fortunately,
one can always use the equation for the total curvature perturbation, $\zeta$ and hence
the problem can be circumvented if only one of the individual curvature perturbations becomes ill-defined.
If more than one component suffers from the same problem, one can use an alternative
formalism, \eg\ by means of individual density perturbations $\delta\rho_{(\alpha)}$, or simply
choose the set of evolving variables suitably before each problematic point.

Here we consider an interacting two-fluid system, where energy density flows from one fluid, $\rho_1$,
to another, $\rho_2$: $Q_1=-\Gamma f(N)\rho_{0(1)},\ Q_2=\Gamma f(N)\rho_{0(1)}$.
During the evolution of the system, the curvature perturbation of the second fluid, $\zeta_{(2)}$ becomes ill-defined at some
point during the period of interest. Therefore, we choose our set of dynamical variables such that the system 
is described by the metric perturbation, $\phi$, the total curvature perturbation, $\zeta$, 
and the curvature perturbation of fluid one, 
$\zeta_{(1)}$. The equations of state of the fluids, $\w_{(i)}$, are taken to be constants
for simplicity. 

In terms of theses variables, the relevant background equations (\ref{fluidcont}) take the form:
\bea{eq:2-fluid-with-deltaQ-N}
\rho_{0}' & = & -3 \left(\rho_{0} + \omega_{(1)}\rho_{0(1)} + \omega_{(2)}(\rho_{0} - \rho_{0(1)}) \right) \\
\rho_{0(1)}' & = & -\left(3(1 + \omega_{(1)}) + \frac{\Gamma f(N)}{H}\right) \rho_{0(1)},
\eea
The equations of motion of the perturbations can be derived from the definition of $\zeta_{(\alpha)}$ and Eqs. (\ref{fluidpert}) and (\ref{psieq})
\bea{eqs1}
\phi' & = & \frac{\rho_{0}'}{2\rho_{0}}(\zeta + \phi) -\phi,\label{eqs1:1}\\
\zeta' & = & -\frac{3\rho_{0(1)}'}{\rho_{0}'}(\omega_{(2)} - \omega_{(1)})\left( \zeta -  \zeta_{(1)} \right),
\label{eqs1:2}\\
\zeta_{(1)}' & = &  \frac{\Gamma \rho_{0(1)}}{H\rho_{0(1)}'}
\Big(\frac{f(N)\rho_{0}'}{2\rho_0}
(\zeta - \zeta_{(1)})
 + f'(N)(\phi + \zeta_{(1)})\Big)\label{eqs1:3},
\eea
The function $f(N)\in [0,1]$ describes the (time-dependent) coupling between the two fluid components.
Physically, $f(N)$ sets how quickly the interaction is turned on up to the full rate $\Gamma$. 
The form of the function $f(N)$ has to be determined by microphysical properties of the interacting 
fluids together with necessary information on the evolution of the whole system. For example,
the interaction strength can naturally be temperature dependent or phase transitions can  
significantly change the properties of the interacting system.
To model and estimate the importance of such effects in general, we here adopt a simple Ansatz
for the interaction without addressing the question of its origin.     
From the equations (\ref{eqs1:1})-(\ref{eqs1:3}) we see how a varying coupling, 
$f'(N)\neq 0$, between the fluids leads to an additional coupling of the metric perturbation, $\phi$, 
to the curvature perturbations, leading to new effects.

\section{Time-dependent interaction}

As a first approximation, we choose the transition to be 
proportional to the Heaviside function, \ie $f(N) = \theta(N - N_{*})$, with 
$N_{*}$ being the time when the transition begins and the strength of the interaction is denoted 
by $\Gamma$. This choice has the advantage that it can be solved analytically.
The fluids are allowed to interact without any further approximations, in 
contrast to the sudden decay approximation, where the
energy transfer between fluids occurs instantly when $\Gamma/H$
 reaches a critical value.

With this choice of $f(N)$, we can integrate the equation of motion of $\zeta_{(1)}$, Eq. (\ref{eqs1:3}).
The integration requires some care and the technical details are presented in the Appendix.
The result is
\begin{equation} \label{eq:zurv1-jump2}
\zeta_{(1)+} = \frac{\left(\phi(N_{*}) + \zeta_{(1)-}\right)\ln[\frac{x_{*}}{1+x_{*}}]}{2+x_{*}\ln[\frac{x_{*}}{1+x_{*}}]} + \zeta_{(1)-},
\end{equation}
where
$x_{*}=(3H(N_{*})(1 + \omega_{(1)}))/\Gamma$ and $\zeta_{(1)\pm}\equiv \zeta_{(1)}(N_{*}\pm)$ are the values right before and after the discontinuous jump. 

Physical processes however rarely happen as quickly as the Heaviside function would require. Therefore a steep hyperbolic tangent is a more realistic choice to model $f(N)$. From the point of view of the dynamics of our system \eg Eq. (\ref{eqs1:3}) this means that during transition, when the term proportional to $f'(N)$ dominates, the system is driven towards a equilibrium where $\zeta_{(1)} = -\phi$. Now the magnitude of $\Gamma$ tells if the system will reach the equilibrium. We can now compare the magnitude of the jump from equation (\ref{eq:zurv1-jump2}) to that of $-\phi_{*}$, \ie if $\zeta_{1+}-\zeta_{1-}$ from eq. (\ref{eq:zurv1-jump2}) is larger than $-\phi_{*}$ our jump-formula is longer valid. This comparison leads to a simple equation for $x_{*}$, which can be solved numerically. The result is that eq. (\ref{eq:zurv1-jump2}) can be used when $\Gamma/H(N_{*}) \ll 11.76(1+\omega_{(1)})$. When we increase $\Gamma$ the system reaches the equilibrium and $\zeta_{(1)+} = -\phi(N_{*}) +\zeta_{1-}$.

In order to determine the value of the metric perturbation $\phi$ and the Hubble parameter $H= \sqrt{\rho_{0}}$ at $N = N_{*}$ (where $8\pi G/3=1$) we need to evolve rest of the equations of motion
until the beginning of the transition. 
Before the transition, $\zeta_{(1)}$ is clearly a constant and hence $\zeta_{(1)-} = \zeta_{(1)}(N_{0})$. 
Therefore the equation of motion of $\zeta$ can be separated and integrated:
\begin{equation} \label{eq:zeta-pre-jump}
\zeta(N) = \frac{(1+\omega_{(1)})\rho_{0(1)}(N_{0}) + (1+\omega_{(2)})\rho_{0(2)}(N_{0})}{(1+\omega_{(1)})\rho_{0(1)}(N_{0})e^{3\Delta\omega\Delta N} + (1+\omega_{(2)})\rho_{0(2)}(N_{0})}\Big(\zeta(N_{0})-\zeta_{(1)}(N_{0})\Big) + \zeta_{(1)}(N_{0}),
\end{equation}
where $N_0$ denotes initial time, and we have defined $\Delta N \equiv N-N_0$ and 
$\Delta\omega \equiv \omega_{(2)}-\omega_{(1)}$.

Unfortunately, we are unable to solve for $\phi(N)$ in the general case and hence we have concentrated
on the two special cases where one of the fluids is initially dominating the energy density of the universe.

\subsection{Dominant decaying fluid: $\rho_{0(1)}(N_*)\gg\rho_{0(2)}(N_*)$}

Before the transition different densities evolve according to
\begin{equation} \label{eq:fluid-one}
\begin{aligned}
\rho_{0}(N) & = \rho_{0(1)}(N_{0})e^{-3(1 + \omega_{(1)})\Delta N} + \Big(\rho_{0}(N_{0}) - \rho_{0(1)}(N_{0})\Big)e^{-3(1 + \omega_{(2)})\Delta N},\\
\rho_{0(1)}(N) & = \rho_{0(1)}(N_{0})e^{-3(1 + \omega_{(1)})\Delta N}.\\
\end{aligned}
\end{equation}
Since $\rho_{(0)1}$ dominates when $N\le N_*$,  
equation (\ref{eq:zeta-pre-jump}) implies that 
\begin{equation} \label{eq:2-fluid-zeta-equation1}
\zeta(N) = \left(\zeta(N_{0}) - \zeta_{(1)}(N_{0})\right)e^{-3\Delta\omega\Delta N} 
+ \zeta_{(1)}(N_{0}).
\end{equation}
Inserting this into the equation of 
motion of $\phi$ and solving the resulting first order differential equation, we have
\begin{widetext}
\bea{eq:2-fluid-phi-equation}
\phi(N)&  \simeq &  \frac{3(1 + \omega_{(1)})}{9\omega_{(1)} - 6\omega_{(2)} + 5}
\Big(\zeta_{(1)}(N_{0}) - \zeta(N_{0})\Big)e^{-3(\omega_{(2)} - \omega_{(1)})\Delta N}
 - \frac{3(1 + \omega_{(1)})}{5 + 3\omega_{(1)}}\zeta_{(1)}(N_{0})\nonumber\\
& &  +  \Big[\phi(N_{0}) + \frac{3(1 + \omega_{(1)})}{9\omega_{(1)} - 6\omega_{(2)} + 5}
\Big(\zeta(N_{0}) - \zeta_{(1)}(N_{0})\Big) + \frac{3(1 + \omega_{(1)})}{5 + 3\omega_{(1)}}
\zeta_{(1)}(N_{0}) \Big]e^{-(5+3\omega_{(1)})\Delta N/2}.
\eea
\end{widetext}
Eqs. (\ref{eq:zurv1-jump2}) and (\ref{eq:2-fluid-phi-equation}) along with the fact
$\zeta_{(1)-}=\zeta_{(1)}(N_0)$ tell us the magnitude of the jump in the curvature
perturbation of fluid one, $\zeta_{(1)+}-\zeta_{(1)-}$. These equations reveal that the 
magnitude of the jump depends crucially on the ratio $\Gamma/H_*$, in addition to the initial values.
Note further that
in most cases where 
$\omega_{(2)} - \omega_{(1)}>0$ or the initial state is adiabatic,
 $\phi (N)$ reaches its 
asymptotic value, $3(1 + \omega_{(1)})\zeta_{(1)}(N_{0})/(5 + 3\omega_{(1)})$,
quite rapidly. Interestingly,
non-adiabatic initial states with $\omega_{(2)} - \omega_{(1)}<0$ may lead to exponentially
amplified curvature perturbations.

In addition to the magnitude of the jump in $\zeta_{(1)}$, an interesting quantity is
the final value of the total curvature perturbation, $\zeta(N_\infty)$.
The magnitude of $\zeta$ also makes a (continuous) jump during the transition, but reaches a constant 
value much more quickly than $\zeta_{(1)}$. This fact allows us to derive quite a useful approximation 
for the final value of the curvature perturbation.

As the energy flows quite rapidly into $\rho_{0(2)}$, 
$\zeta$ becomes a constant almost immediately after the transition. 
In addition, since $\rho_{0(1)}$ vanishes quickly, $\zeta$
changes more rapidly than $\zeta_{(1)}$ and hence during this time we
can approximate $\zeta_{(1)} \approx \zeta_{(1)+}$. 
Similarly, 
since $H'(N)/H(N) = -3(1 + \omega)/2$, with 
$\omega \equiv (\omega_{(1)}\rho_{0(1)} + \omega_{(2)}\rho_{0(2)})/\rho_{0}$ being the effective
equation of state, 
the Hubble parameter changes slowly compared 
to $\rho_{0(1)}$ and hence $H(N) \simeq H(N_{*}) \equiv H_{*}$
during the transition.

Based on these arguments we can estimate the difference between the relative rates of change of 
$\rho_{0(1)}$ and $\rho_{0}$:
\begin{equation} \label{eq:rho1-dom-est-1}
\frac{\rho_{0(1)}'}{\rho_{0(1)}} - \frac{\rho_{0}'}{\rho_{0}} = -3(1 + \omega_{(1)}) - \frac{\Gamma}{H(N)} + 3(1+\omega) \simeq -\frac{\Gamma}{H_{*}} + 3\Delta\omega_{*},
\end{equation}
where $\Delta\omega_{*} = \omega_{*} - \omega_{(1)}$ and $\omega_{*}\equiv \omega(N_{*})$.
Assuming that $\Gamma/H_{*} > 3\Delta\omega_{*} $ and integrating both sides of equation (\ref{eq:rho1-dom-est-1}), we obtain
\begin{equation}
\rho_{0(1)}(N) \simeq \frac{\rho_{0(1)}(N_{*})}{\rho_{0}(N_{*})}\exp[(3\Delta\omega_{*} -\frac{\Gamma}{H_{*}})(N-N_{*})]\rho_{0}(N).
\end{equation} 
Using this in the  
equation of motion of $\zeta$, Eq. (\ref{eqs1:2}),
we get
\begin{equation} \label{eq:zeta-final-1-dom}
\zeta(N_{\infty}) = \zeta_{(1)+} - \Big(\zeta_{(1)+} - \zeta_{*}\Big)\Big[\frac{-\Delta\omega\rho_{1*}+(1+\omega_{(2)})\rho_{*}}{(1+\omega_{(2)})\rho_{*}}\Big]\exp[\frac{3\Delta\omega}{3\Delta\omega_{*} - \frac{\Gamma}{H_{*}}}\frac{\rho_{1*}}{\rho_{*}}],
\end{equation}
where $\zeta_{*} = \zeta(N_{*})$, $\rho_{1*} = \rho_{0(1)}(N_{*})$ and $\rho_{*} = \rho_{0}(N_{*})$. 
When the system in initially dominated by decaying fluid, we may set $\omega_* \simeq \omega_1$. 
This equation along with Eqs. (\ref{eq:zurv1-jump2}) and (\ref{eq:zeta-pre-jump})
determine the final value of the total curvature perturbation
in terms of initial conditions and interaction strength ratio $\Gamma /H_*$.

\subsection{Sub-dominant decaying fluid: $\rho_{(0)1}(N_*)\ll \rho_{(0)2}(N_*)$}

Supposing that the  second fluid dominates before transition, {\it i.e.} $\rho_{0(1)} \ll \rho_{0(2)}$ we have
\begin{equation} \label{eq:fluid-two}
\rho_{0}(N) \simeq \rho_{0(2)}(N) = \rho_{0(2)}(N_{0})e^{-3(1 + \omega_{(2)})\Delta N}.
\end{equation}
Hence it follows from equation (\ref{eq:zeta-pre-jump}), that
\begin{equation}
\zeta(N) = \zeta(N_{0}),
\end{equation}
regardless of the initial conditions of $\zeta$ and $\zeta_{(1)}$. The corresponding solution of $\phi$ is
\begin{equation}\label{eqphi2}
\phi(N) \simeq \Big(\phi(N_{0}) + \frac{3(1+\omega_{(2)})\zeta({N_{0}})}{3\omega_{(2)}+5}\Big)e^{-(3\omega_{(2)}+5)\Delta N/2} - \frac{3(1+\omega_{(2)})\zeta({N_{0}})}{3\omega_{(2)}+5}.
\end{equation}
From this we can determine $\phi(N_*)$ and hence $\zeta(N_\infty)$ using equation (\ref{eq:zeta-final-1-dom}), since $\omega_*\simeq \omega_2$. Note that equation (\ref{eqphi2}) works well only when the system is adiabatic. Despite this our equation for the final value of total curvature perturbation \ie $\zeta(N_{\infty})$ gives accurate values as can be seen from Fig. \ref{fig3c}.

\subsection{Numerical results}

\begin{figure}[tbh]

\subfigure[\ The evolution of a matter-radiation system.]{\label{fig1a}\includegraphics[width=0.47\columnwidth]{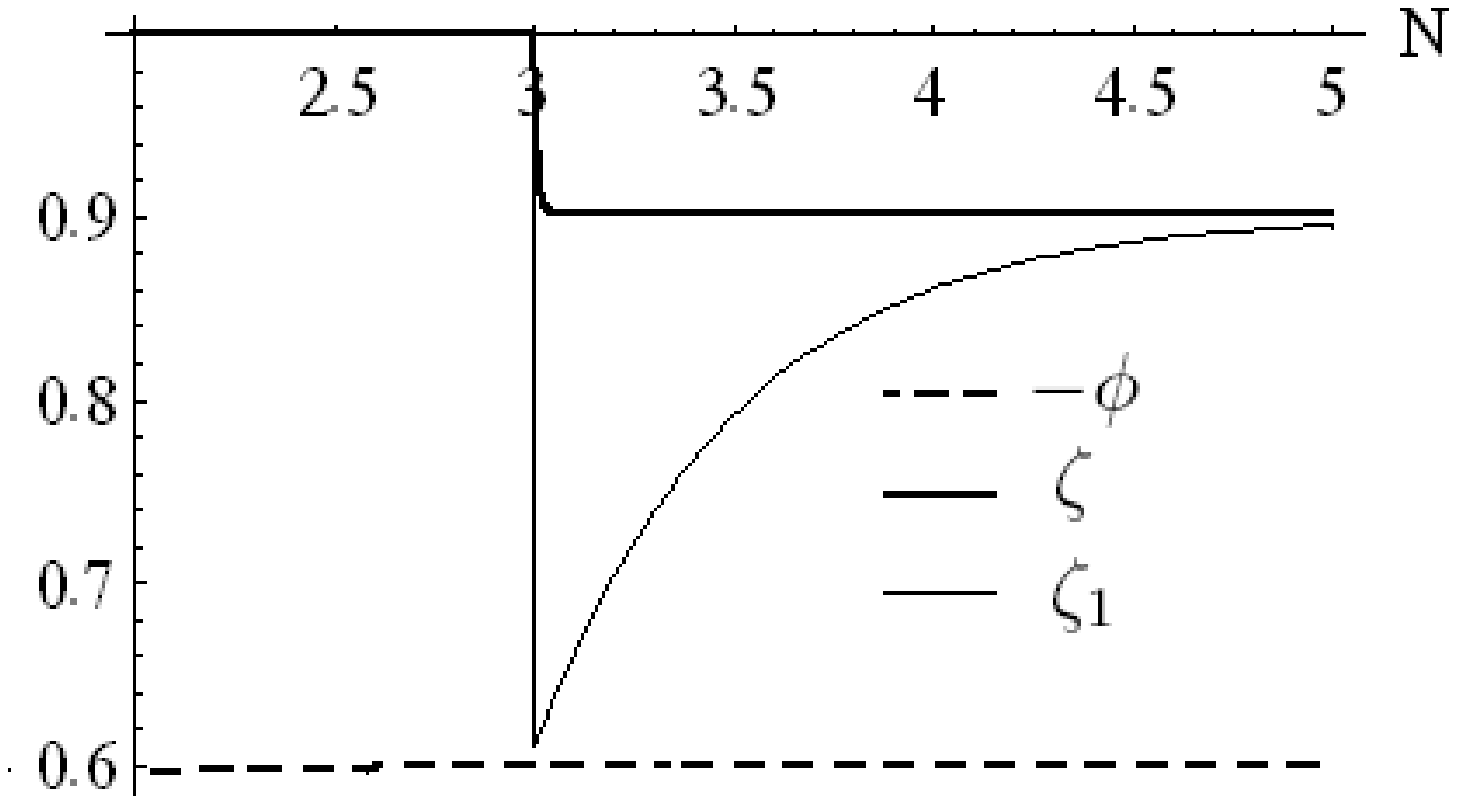}}
\subfigure[\ Ratio of analytical to numerical values of $\zeta_{(1)+}$.]{\label{fig1b}\includegraphics[width=0.50\columnwidth]{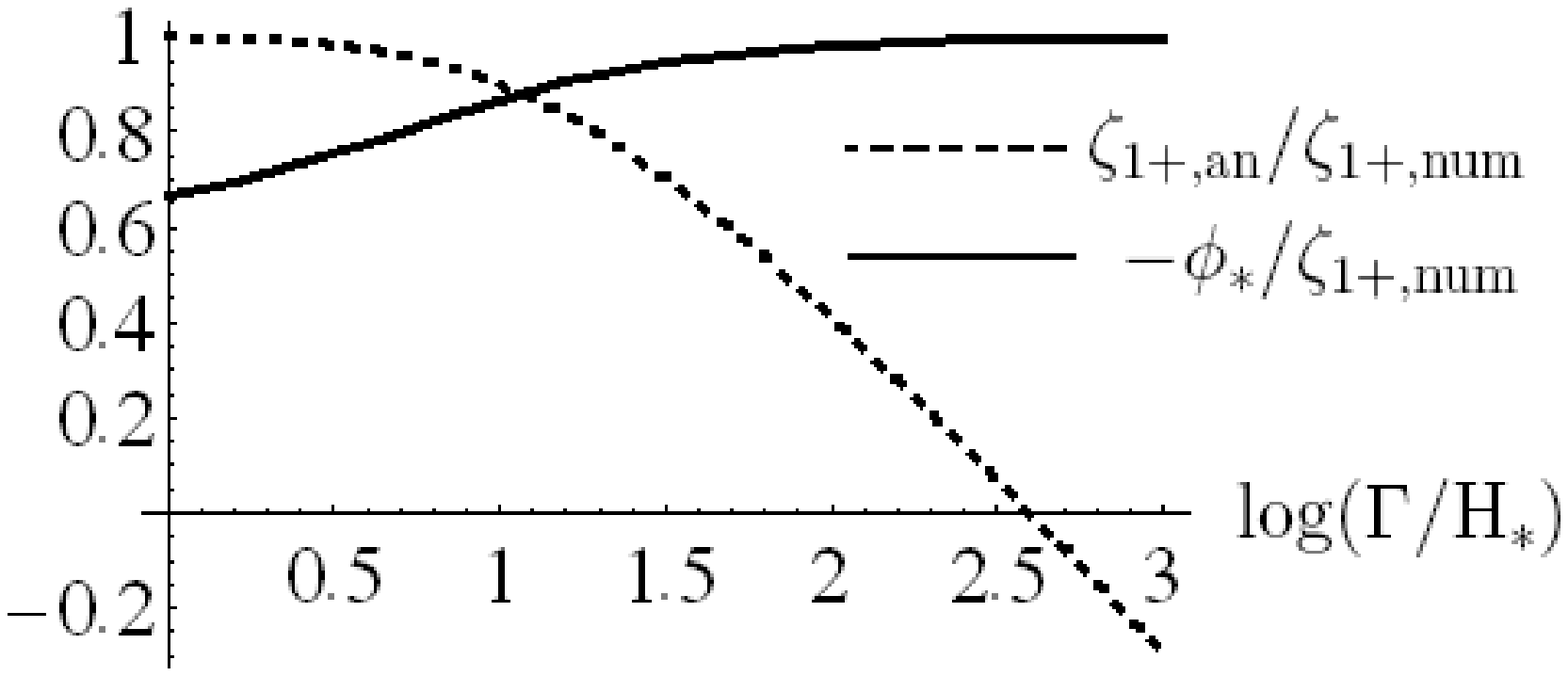}}
\caption{An example of a matter-radiation system with initial matter domination (a). 
Comparison between the numerical and analytical values of the jump $\zeta_{(1)+}$ as a function $\tau$ for different values of the interaction strength, $\Gamma/H_{*}$ (b).}
\end{figure}

\begin{figure}[tbh]
\subfigure[\ Non-adiabatic matter domination initially.]{\label{fig3a}\includegraphics[width=0.47\columnwidth]{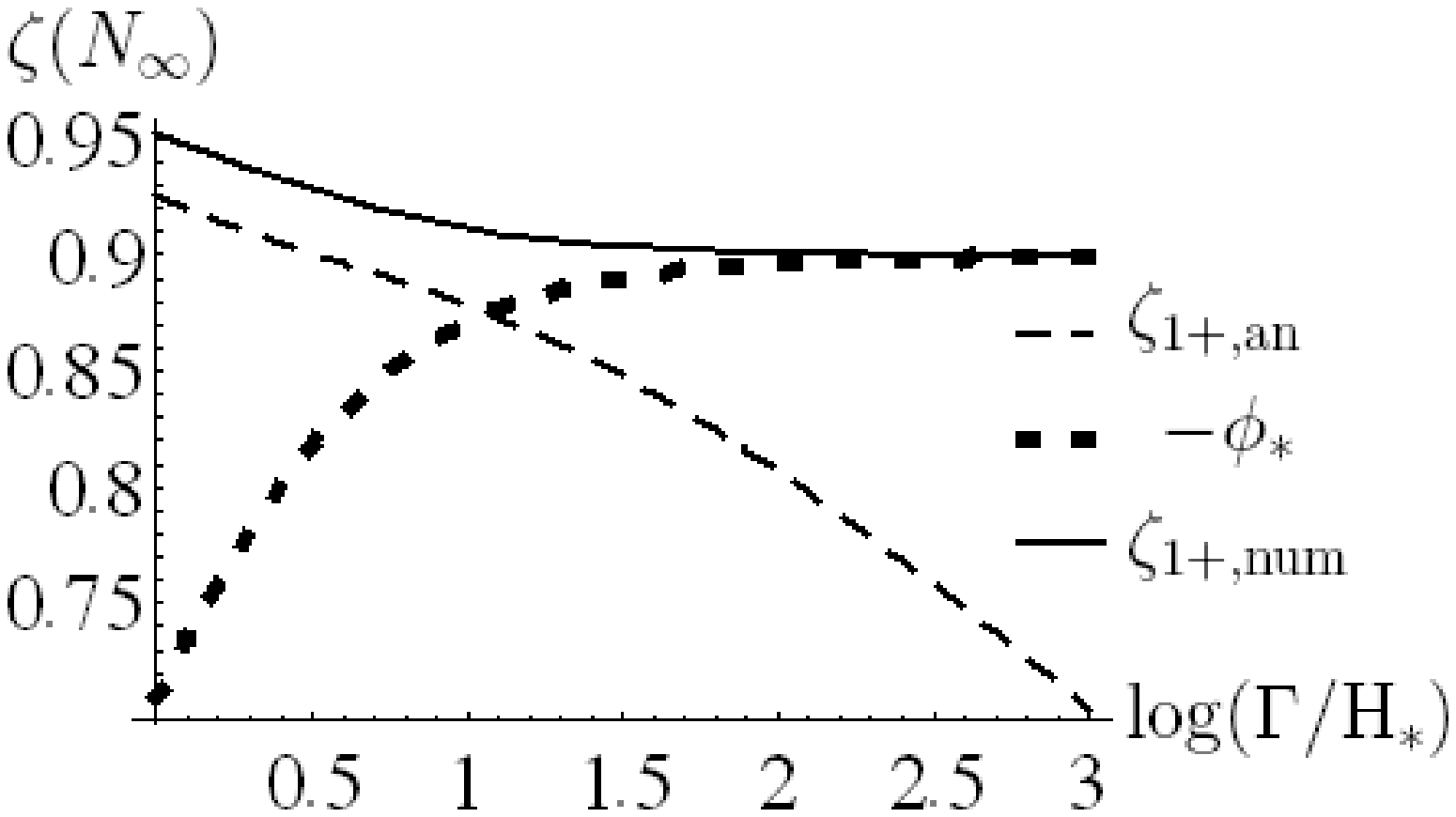}}
\subfigure[\ Adiabatic radiation domination initially.]{\label{fig3b}\includegraphics[width=0.47\columnwidth]{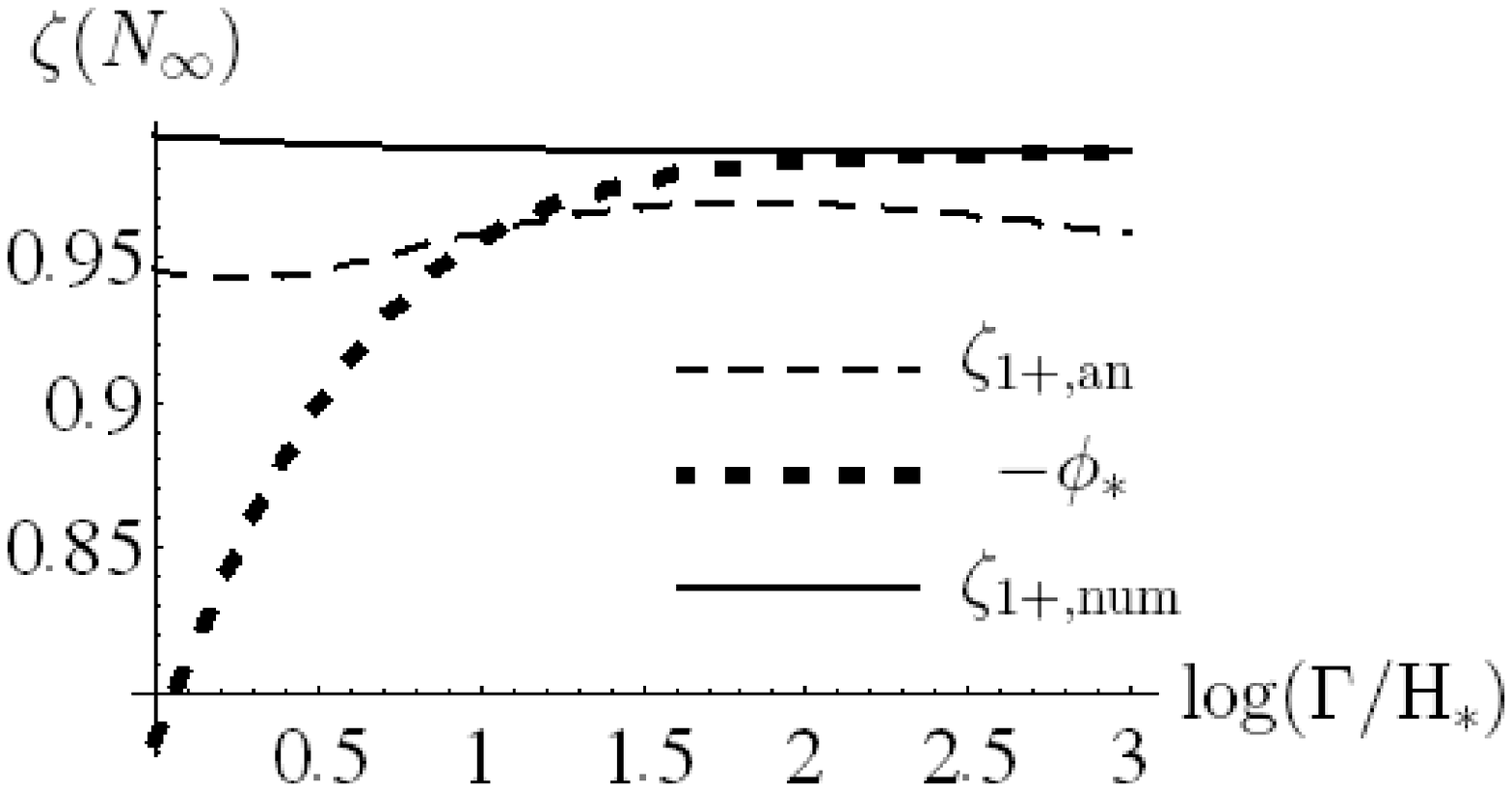}}
\subfigure[\ Non-adiabatic radiation domination initially.]{\label{fig3c}\includegraphics[width=0.47\columnwidth]{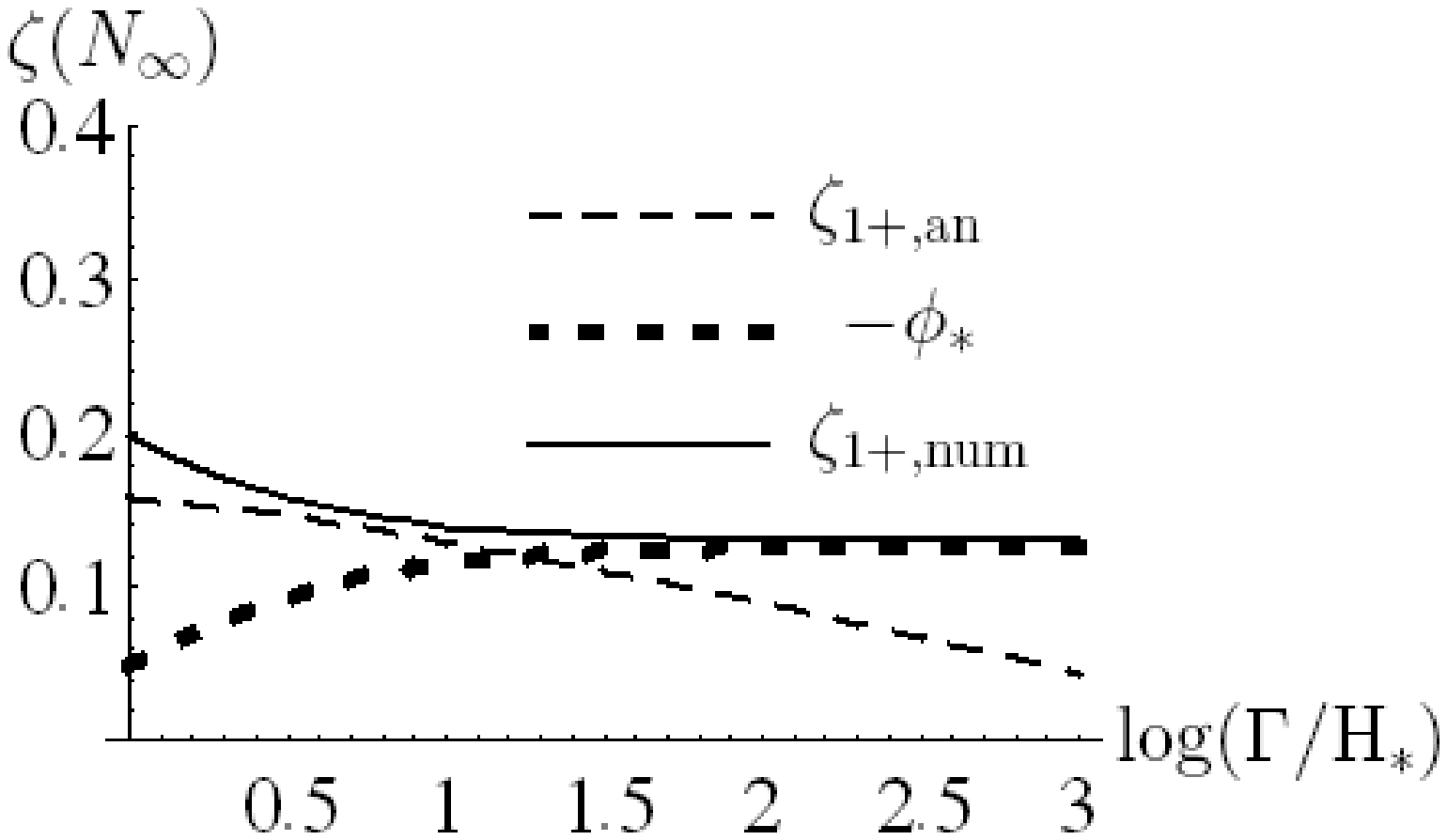}}
\caption{\subref{fig3a} The comparison between numerical (continuous) and different analytical values (dashed)
 of $\zeta(N_\infty)$, in a matter-radiation system when the initial state is 
dominated by matter (a), adiabatic radiation (b) and non-adiabatic radiation (c).}
\end{figure}

In order to study the applicability of these results, we have compared the analytical values with those
obtained from full numerical evolution of the equations. An example of the numerical calculations 
is shown in  Fig. \subref{fig1a} where we have calculated the evolution of a system where
an initially dominating  matter fluid ($\omega_{(1)}=0$) that decays into radiation ($\w_{(2)}=1/3$).
Such a transition occurs \eg when the inflaton decays to radiation via coherent oscillations or in
the curvaton scenario (although it is typically assumed that the curvaton is initially subdominant, see \eg \cite{Malik:2002jb}).
The initial values of the energy densities are 
$\rho_{(1)}(N_{0}) = 0.999$, $\rho(N_{0}) = 1.0$ and the initial state is adiabatic, 
$\zeta(N_{0}) = \zeta_{(1)}(N_{0}) = 1.0$ with $\phi(N_0) = 0.01$.

The interaction strength ratio is given by
$\Gamma/H_* = 100$ and the transition function by 
$f(N)= (\tanh[(N-N_{*})/\tau]+1)/2$. In order to model the rapid transition we 
choose $\tau = 1.0 \times 10^{-5}$ with $N_{*}-N_0=3$.

Figure \subref{fig1a} shows how $\zeta_{(1)}$ experiences a sudden decrease at $N=3$, until it slowly begins to decay. Similarly, the total curvature perturbation, $\zeta$, begins to decline rapidly after the jump until
it quickly reaches a constant value. The evolution of the metric perturbation, $\phi$, is less dramatic
and it reaches a constant value at a rate similar to that of $\zeta_{(1)}$. The final state is
again adiabatic, $\zeta=\zeta_{(1)}$, as expected.

If the second fluid dominates from the beginning, like in the curvaton scenario, the behavior 
of perturbations is quite similar to the one pictured above: the jump of $\zeta_{(1)}$ depends 
mainly on the magnitude of $\Gamma/H_{*}$ and the initial values. 
The final value of $\zeta$ again depends on $\Gamma/H_{*}$, but it is more strongly 
related to the initial values of $\rho_{0(1)}$ and $\rho_{0}$.

The comparison between the analytical result, Eq. (\ref{eq:zurv1-jump2}), and the numerical calculation
is shown in \subref{fig1b}. In the figure \subref{fig1b} we plot the ratio of the analytical formula to the numerical result
for different values of $\Gamma/H_{*}$. As can be seen from the figure, the analytical 
formula, Eq. (\ref{eq:zurv1-jump2}), is very good when $\Gamma/H_* < 1$ but it
begins to over-estimate the magnitude for larger $\Gamma/H_*$. For $\Gamma/H_*> 11.75$ the value of $\zeta_{(1)+} = -\phi(N_{*})$ 
is a better estimate.

The comparisons between the full numerical calculation and the analytical approximation for asymptotic values,
Eq. (\ref{eq:zeta-final-1-dom}), are presented in Fig. 2 for the same system as studied above with
the first fluid initially dominant (a) or sub-dominant (b).
In the first case the initial densities of fluids have been set to $\rho_{(1)}(N_{0}) = 0.99$, $\rho(N_{0}) = 1.0$ and the system has been chosen to be non-adiabatic. However, since fluid one dominates the system is very close to the adiabatic state.
In the second case where the second fluid is dominating from the beginning, we set
$\rho_{(1)}(N_{0}) = 10^{-2.0}$ and $\rho(N_{0}) = 1.0$. We have calculated the evolution of the system with adiabatic and non-adiabatic initial perturbations. The latter of these is related to the curvaton scenario, where the initial values are $\zeta_{(1)}(N_{0}) = 1.0$, $\zeta_{(2)}(N_{0}) = 0$. The metric perturbation is chosen as in the previous case. From the figure we see that the analytical formulae generally give a good approximation to the exact numerical calculation.
Again, when $\Gamma/H_{*}$ is smaller, the approximation  given by Eqs. (\ref{eq:zurv1-jump2}) and (\ref{eq:zeta-final-1-dom}) is better, which is what we expect since then $f(N)$ is more closely approximated by the Heaviside function.

\section{Conclusions and discussion}

In the present article we have discussed the evolution of perturbation variables
in a two-component interacting fluid system. This has been studied previously
by a number of authors, \eg \cite{Kodama:1985bj, Wands:2000dp, Malik:2004tf, Matarrese:2003tk}. 
Our treatment goes beyond the sudden decay approximation by 
including all relevant perturbations explicitly and including a time scale, $\tau$,
that describes how quickly the interaction is turned on. Such an interaction is well motivated
by physical considerations and worth considering particularly in the fluid approach, where
the microphysical properties of the fluids are otherwise hidden.

We have seen how the the curvature perturbation can be efficiently suppressed
in a system with interacting fluids. A fluid that decays rapidly, $\tau H_* \ll 1$, $\Gamma/H_* \gg 1$,
generally diminishes the curvature perturbation associated with that fluid which then
subsequently affects the total curvature perturbation.
The analysis shows that the decrease of perturbations depends crucially on 
the ratio $\Gamma/H_*$ in addition to the initial conditions. 

We have derived two expressions for different values of  $\Gamma/H_*$ which give the magnitude 
of this jump and compared these analytical results to the numerical calculations, showing good agreement.
We have also derived a formula to approximate the final value of the total curvature perturbation, 
$\zeta$, after the transition. Again we find that the approximation is good when the decay is 
fast enough, \ie $\Gamma/H_*\gsim 10^{-1}$.

Processes with efficient suppression of the large-scale curvature perturbation  
can in general be realized in a scenario
where the three physical time scales have a hierarchy, $H_* \ll \Gamma,\ 1/\tau$.
Such a hierarchy can be achieved naturally if the time scales are associated with different physical 
processes. If, on the other hand, the decay processes become effective only when $\Gamma\sim H_*$, 
no large net effect results.
Interesting avenues to explore in this respect are processes including phase transitions 
and/or different scales, making  many cosmological phase transitions a potential source of 
interesting phenomena.  In addition, inflationary scenarios with multiple phases can provide a suitable
framework for modifying the large-scale perturbations, \eg  when the
intermediate phases between inflationary periods are dominated by 
string networks \cite{cliff}.

\subsection*{Acknowledgments}
This work has been partly funded by the Academy of Finland, project no. 8111953.
TM and JS are supported by the Academy of Finland.


\appendix*\section{}
Naive integration of the equation of motion for
$\zeta_{(1)}$ (\ref{eqs1:3}) 
over the interval $[N_{*} - \epsilon, N_{*} + \epsilon]$, where $\epsilon > 0$, gives
\begin{widetext}
\bea{eq:2-fluid-jump-int-1}
\int^{N_{*} + \epsilon}_{N_{*} - \epsilon} \zeta_{(1)}' dN & = & \zeta_{(1)}(N_{*} + \epsilon) - \zeta_{(1)}(N_{*} 
- \epsilon)\nonumber\\
& = &  \int^{N_{*} + \epsilon}_{N_{*} - \epsilon} \Big(\frac{\Gamma \theta(N - N_{*}) \rho_{0(1)}\rho_{0}'}
{2H\rho_{0}\rho_{0(1)}'}(\zeta - \zeta_{(1)}) 
 + \frac{ \rho_{0(1)}\Gamma \theta'(N - N_{*})}{H\rho_{0(1)}'}(\phi + \zeta_{(1)})\Big) dN.
\eea
\end{widetext}
By going to the limit $\epsilon \rightarrow 0+$, the term proportional to $\zeta - \zeta_{(1)}$
clearly disappears, since the integrand is limited and the interval of integration goes to zero. 
The other term requires more care. Clearly 
the numerator of the integrand is proportional to $\theta'(N - N_{*})$
(or $\delta(N - N_{*})$) while the denominator has a term proportional to the $\theta(N-N_*)$.
In order to handle this term properly, we include a $C^\infty$ test function 
(with compact support) $\varphi(N)$ into our integrals and handle the integrals in terms of weak convergence of
distributions. 

We study integrated form of Eq. (\ref{eqs1:3})
\be{A1}
 \int^{N_{*} + \epsilon}_{N_{*} - \epsilon}\zeta_{(1)}'(N) \varphi(N)\, dN =   \int^{N_{*} + \epsilon}_{N_{*} - \epsilon}
 \frac{\Gamma \rho_{0(1)}}{H\rho_{0(1)}'}
\Big(\frac{f(N)\rho_{0}'}{2\rho_0}
(\zeta(N) - \zeta_{(1)}(N))
 + f'(N)(\phi(N) + \zeta_{(1)}(N))\Big)\varphi(N)\, dN
\ee
and after evaluations take the limit $\epsilon \rightarrow 0+$.
We also split the $\zeta_{(1)}(N)$-function into two parts: the continuous
part, $\bar{\zeta}_{(1)}(N)$,  and the jump $\Delta\zeta_{(1)}\theta(N - N_{*})$, where 
$\Delta\zeta_{(1)} =\zeta_{(1)}(N_{*}+)-\zeta_{(1)}(N_{*}-)\equiv  \zeta_{(1)+} - \zeta_{(1)-}$. 

From l.h.s.\ of Eq. (\ref{A1}) we get
\begin{equation}
I_{1}(N_{*}) \equiv \lim_{\epsilon \rightarrow 0+} \int^{N_{*} + \epsilon}_{N_{*} - \epsilon}\zeta_{(1)}'\varphi(N)dN 
= \Delta\zeta_{(1)}\varphi(N_{*}).\\
\end{equation}
and first term on r.h.s. vanish at $\epsilon$-limit. 
Next we take the the continuous part of 
$\phi + \zeta_{(1)}$ term on the right side of equation (\ref{A1}) which yields
\begin{widetext}
\begin{equation} \label{eq:I2}
\begin{aligned}
I_{2}(N_{*}) \equiv & -\lim_{\epsilon \rightarrow 0+}\int^{N_{*} + \epsilon}_{N_{*} - \epsilon}\varphi(N)\left(
\frac{\theta'(N-N_{*})}{x(N) + \theta(N-N_{*})}(\phi + \bar{\zeta}_{(1)})\right)dN \\
= & -\lim_{\epsilon \rightarrow 0+}\int^{\infty}_{-\infty}\varphi(N)\left(\chi_{[N_{*} - \epsilon,N_{*} + \epsilon]}
\frac{d}{dN}\ln[x(N) + \theta(N - N_{*})]\right)(\phi + \bar{\zeta}_{(1)})dN\\
& + \lim_{\epsilon \rightarrow 0+}\int^{N_{*} + \epsilon}_{N_{*} - \epsilon}\varphi(N)\left(\frac{x'(N)}{x(N) + 
\theta(N - N_{*})}\right)(\phi + \bar{\zeta}_{(1)})dN,\\
\end{aligned}
\end{equation}
\end{widetext}
where we have introduced the function $\chi_{[N_{*} - \epsilon,N_{*} + \epsilon]}$,
\begin{equation}
\chi_{[N_{*} - \epsilon,N_{*} + \epsilon]} = \theta(N - N_{*} + \epsilon) - \theta(N - N_{*} - \epsilon),
\end{equation}
and defined $x(N) = 3H(N)(1+\omega_{(1)})/\Gamma$.

The last term of (\ref{eq:I2}) vanish at the limit $\epsilon\rightarrow 0+$ and we can now integrate  
the other term in parts when all terms not proportional 
to the derivative of $\chi_{[N_{*} - \epsilon,N_{*} + \epsilon]}$ gives
vanishing contribution to the integral in the limit $\epsilon \rightarrow 0+$, too.
Thus,
\begin{equation}
\begin{aligned}
I_{2}(N_{*}) & = \varphi(N_{*})(\ln[x_{*}] - \ln[x_{*} + 1])(\phi_{*} + \bar{\zeta}_{(1)*}).\\
\end{aligned}
\end{equation}

Finally, we integrate the jump-part of $\phi+\zeta_{(1)}$ term:
\bea{A7}
I_{3}(N_{*}) &=& -\lim_{\epsilon \rightarrow 0+} \int^{N_{*} + \epsilon}_{N_{*} - \epsilon}
\Delta\zeta_{(1)}\varphi(N)\theta'(N-N_{*})dN \nonumber + \lim_{\epsilon \rightarrow 0+} \int^{N_{*} + \epsilon}_{N_{*} - \epsilon} 
\frac{\Delta\zeta_{(1)}\varphi(N)x(N)\theta'(N-N_{*})}{x(N) + \theta(N-N_{*})}dN
\eea
The first term in r.h.s.\ is clearly $\Delta\zeta_{(1)}\varphi(N_{*})$ and the second term 
can be written in the form
\begin{widetext}
\begin{equation}
\begin{aligned}
I_{3b}(N_{*}) \equiv &  \lim_{\epsilon \rightarrow 0+} \int^{\infty}_{-\infty} \Big\{\Delta\zeta_{(1)} x(N)
\varphi(N)\chi_{[N_{*} - \epsilon,N_{*} + \epsilon]} \left(\frac{d}{dN}\ln[x(N) + \theta(N - N_{*})] - 
\frac{x'(N)}{x(N) + \theta(N-N_{*})}\right)\Big\}dN.\\
\end{aligned}
\end{equation}
\end{widetext}
Again term proportional to $x'(N_{*})$ vanishes when $\epsilon \rightarrow 0+$
and the other term can be integrated by part. We obtain 
\begin{equation}
\begin{aligned}
& I_{3}(N_{*}) =& -\Delta\zeta_{(1)}\varphi(N_{*}) - \Delta\zeta_{(1)}x_{*}\varphi(N_{*})\left(\ln[x_{*}] - \ln[x_{*} + 1]\right).\\
\end{aligned}
\end{equation}

Now, because Eq.\ (\ref{A1}) implies
\be{A2}
 I_1=I_2+I_3,
\ee
by adding up all the terms and canceling all non-zero common factors  we find
\begin{equation} \label{eq:zurv1-jump}
\zeta_{(1)+} = \frac{\left(\phi(N_{*}) + \zeta_{(1)-}\right)\ln[{\frac{x_{*}}{1+x_{*}}}]}{2+x_{*}\ln[\frac{x_{*}}{1+x_{*}}]} + \zeta_{(1)-},
\end{equation}
where the parameter $x_*$ is defined by
\begin{equation}
x_{*}\equiv \frac{3H(N_{*})(1 + \omega_{(1)})}{\Gamma}.
\end{equation}


\end{document}